\documentclass[aps,prl,twocolumn,showpacs,epsfig,floats,superscriptaddress]{revtex4}
\usepackage{amsmath}
\usepackage{color}
\usepackage{graphicx}
\usepackage{amsfonts}
\usepackage{mathrsfs}
\usepackage{amsmath}

\begin{document}
\title{\textbf{Signature of Chaos and Delocalization in a Periodically Driven Many Body System : An Out-of-Time-Order Correlation Study}}

\author{S. Ray}
\affiliation{Indian Institute of Science Education and Research,
Kolkata, Mohanpur, Nadia 741246, India}

\author{S. Sinha}
\affiliation{Indian Institute of Science Education and Research,
Kolkata, Mohanpur, Nadia 741246, India}

\author{K. Sengupta}
\affiliation{Theoretical Physics Department, Indian Association for
the Cultivation of Science, Jadavpur, Kolkata-700032, India.}

\date{\today}
\begin{abstract}

We study out-of-time-order correlation (OTOC) for one-dimensional
periodically driven hardcore bosons in the presence of Aubry-Andr\'e
(AA) potential and show that both the spectral properties and the
saturation values of OTOC in the steady state of these driven
systems provide a clear distinction between the localized and
delocalized phases of these models. Our results, obtained via exact
numerical diagonalization of these boson chains, thus indicate that
OTOC can provide a signature of drive induced delocalization even
for systems which do not have a well defined semiclassical (and/or
large $N$) limit. We demonstrate the presence of such signature by
analyzing two different drive protocols for hardcore bosons chains
leading to distinct physical phenomena and discuss experiments which
can test our theory.

\end{abstract}

%\date{}
\maketitle

{\it Introduction :} Identifying the signature of chaos in quantum
systems is a long standing issue \cite{Haake_b,Casati} which has
relevance for both its entanglement properties \cite{Ent_exp,Greiner} and
thermalization \cite{chaosref,chaosref1}. Typical fingerprint of
chaos in a quantum system may be found in its spectral properties by
invoking the BGS conjecture \cite{Bohigas}. Recent studies however
show that the out-of-time-order correlator (OTOC) provides an
alternate and more direct way to quantify chaos even in the
interacting many body systems
\cite{russian1,Yoshii,Swingle1,Demler,Knap,Galitski1,Galitski,Fradkin,Heyl,Bordia}.
Recent developments in the experimental techniques to measure the
quantum correlations enables a direct investigation of the OTOC in
trapped ions and spin systems \cite{ARey,OTOC_exp}. For quantum
systems with a well-defined semiclassical limit, OTOC provides a way
to estimate the Lyapunov exponent which may be used to quantify the
degree of chaos of the system \cite{russian1}. Interestingly
application of this method in the Sachdev-Ye-Kitaev (SYK) model
\cite{Sachdev1} provides an upper bound to this Lyapunov exponent
which is believed to have a connection with the information
scrambling in black holes \cite{blackhole1}. For the same reason
this method has its application in quantum information as well as in
study of the entanglement in strongly interacting quantum systems
\cite{Ent_th,Swingle2}.

On the other hand, study of periodically driven many-body systems
has regained interest after the recent experimental observation of
drive induced delocalization phenomena \cite{Bloch}. The study of an
equivalent non-interacting model reveals that such delocalization
phenomena stems from the underlying chaotic dynamics \cite{SRay}. In
this context OTOC turns out to be an ideal method to explore the
connection between the delocalization and the underlying chaos in an
interacting quantum system. Although the connection of OTOC with
Lyapunov exponent has been explored in several condensed matter
systems \cite{Swingle1,Demler,Knap,Galitski1,Galitski}, to the best
our knowledge the delocalization transition from many
body localized (MBL) phases of quasiperiodic systems has not been
investigated so far using OTOC.

The experimental realization of quasiperiodic system such as the
Aubry-Andr\'e (AA) model has become a testbed to study single
particle \cite{Inguscio} as well as MBL phenomena of strongly
interacting systems \cite{Bloch_exp} since the AA model exhibits
localization transition in 1D \cite{Aubry,Hanggi}. In a recent
experiment the dynamics of many body localized two component
fermions subjected to a driven AA model reveals delocalization
phenomena controlled by the frequency of the drive \cite{Bloch}.
Motivated by this experiment, in this work we consider a system of
strongly interacting bosons in the presence of an AA potential
subjected to two different types of periodic drives which have
different consequence on delocalization phenomena. Our goal is to
study the commutator
\begin{eqnarray}
C(\beta_{T},p) = Tr\left[\hat{\rho}_{\beta_{T}}
[\hat{W}(p),\hat{V}(0)]^{\dagger}[\hat{W}(p),\hat{V}(0)]\right]
\label{correq1}
\end{eqnarray}
calculated after $p^{\rm th}$ drive cycle using the thermal density
matrix $\hat{\rho}_{\beta_{T}}$at inverse temperature
$k_B\beta_{T}$ (where $k_B$ is the Boltzmann constant), of suitable
local unitary operators $\hat{W}(p) \equiv \hat W(t=pT)$ and
$\hat{V}$ and to detect the delocalization transition in these
driven systems from its behavior. We note that
$C(\beta_T, p)$ is related to the OTOC defined as
\begin{equation}
F(\beta_T,p) = Tr \left[\hat{\rho}_{\beta_T}
\hat{W}^{\dagger}(p)\hat{V}^{\dagger}(0)\hat{W}(p)\hat{V}(0) \right]
\label{f}
\end{equation}
via $C(\beta_T,p) = 2(1 - Re[F(\beta_T,p)])$. The
last relation holds for operators $\hat{W}$ and $\hat{V}$ that
satisfy $\hat{W}^2=\hat{V}^2=\hat{I}$; we shall always focus on such
operators here.

Since in the semiclassical limit the Lyapunov exponent of the
corresponding quantum system can be estimated from the unequal time
commutator of conjugate dynamical variables \cite{Galitski1}, it is
natural to expect that $F(\beta_T,p)$ (and equivalently
$C(\beta_T,p)$) would capture thermalization and underlying chaos in
a quantum many body system and thereby distinguish
between its MBL and ergodic phases. In this work, by carrying out a
detailed study of properties of the OTOC for two periodically driven
boson models in the presence of an Aubry-Andr\'e (AA) potential, we
show that that this is indeed the case. We discuss the dependence of
the OTOC on the time period $T$ of the drive and show that both its
saturation value and spectral properties can distinguish between MBL
and ergodic phases; our results thus show that these quantities can
serve as an indicator of delocalization transition even for systems
with no obvious semiclassical or large $N$ limit and thus with no
clear definition of Lyapunov exponents.

We start by constructing the Floquet operator to generalize OTOC for
stroboscopic dynamics and compare the behavior of $C(\beta_T,p)$
with the spectral properties of the Floquet operator which has
become a standard method to identify delocalization transition. The
most general Hamiltonian describing a system under periodic
perturbation is given by,
\begin{equation}
\hat{H}(t) = \hat{H}_{0} + \hat{H}_{1}(t)
\label{ham_t}
\end{equation}
where $\hat{H}_{0}$ is time independent part and the time dependent
part satisfies $\hat{H}_{1}(t + T) = \hat{H}_{1}(t)$, where $T$ is
the time period of the drive. The corresponding Floquet operator is
$\hat{\mathcal{F}} = \hat{\mathcal{T}} e^{-i \int _{0}^{T}
\hat{H}(t) dt /\hbar}$, where $\hat{\mathcal{T}}$ is
the time ordering operator. Due to unitarity of $\mathcal{\hat{F}}$,
the eigenvalue equation can be written as :
$\mathcal{\hat{F}}|\psi_{\nu}\rangle = e^{-i \phi_{\nu}}
|\psi_{\nu}\rangle$, where $\phi_{\nu}$ and $|\psi_{\nu}\rangle$ are
the eigenphase and the eigenstate corresponding to the $\nu$th
eigenmode of $\hat{\mathcal{F}}$. For periodically driven system the
OTOC, $F$ corresponding to two unitary operators $\hat{W}$ and
$\hat{V}$ after the $p^{\rm th}$ drive cycle is given by Eq.\
\ref{f} with $\hat{W}(p) = \mathcal{\hat{F}}^{\dagger p} \hat{W}(0)
\mathcal{\hat{F}}^{p}$. In what follows, we shall describe two
physical models and analyze the effect of the drive from the
properties of Floquet operator and OTOC.

{\it Model I :} We consider a periodically driven system of hardcore
bosons within tight binding approximation given by the Hamiltonian,
\begin{subequations}
\begin{eqnarray}
\hat{H}_0 &=& -J\sum_{l} \left(\hat{b}_{l}^{\dagger}\hat{b}_{l+1}+h.c.\right) + \mathcal{V} \sum _{l} \hat{n}_l \hat{n}_{l+1} \\
\hat{H}_{1}(t) &=& \lambda \left(1+\epsilon f(t)\right)
\sum_{l}\cos(2\pi \beta l)\hat{n}_{l},
\end{eqnarray}
\label{ham1}
\end{subequations}
where $\hat{b}_{l}^{\dagger}$ and $\hat{n}_{l} =
\hat{b}_{l}^{\dagger} \hat{b}_{l}$ are the creation and the density
operators of the bosons at the $l^{\rm th}$ lattice site respectively, $J$
is the hopping amplitude, $\mathcal{V}$ is the strength of the
nearest neighbor interaction, $\lambda$  denotes the amplitude of the
quasiperiodic potential, and $\beta = (\sqrt{5}-1)/2$. For simplicity we consider
a square pulse protocol in the interval $x= 2 \pi t/T \in [0,2 \pi]$
given by $f(x) = \theta (x - \pi) - \theta (\pi - x)$, where
$\theta(x)$ is the Heaviside step function. In rest of the paper, we set
$\hbar = 1$, all energies (times) are measured in unit of $J(1/J)$ and we
consider $\lambda = 3$ and $\epsilon = 0.47$ such that the time
independent Hamiltonian represents the localized regime of the AA
model and drive induces mixing with the delocalized regime. All the
plots shown below are done for number of lattice sites $N_s = 12$ at
the half filling.

\begin{figure}[ht]
\centering
\includegraphics[scale=0.16]{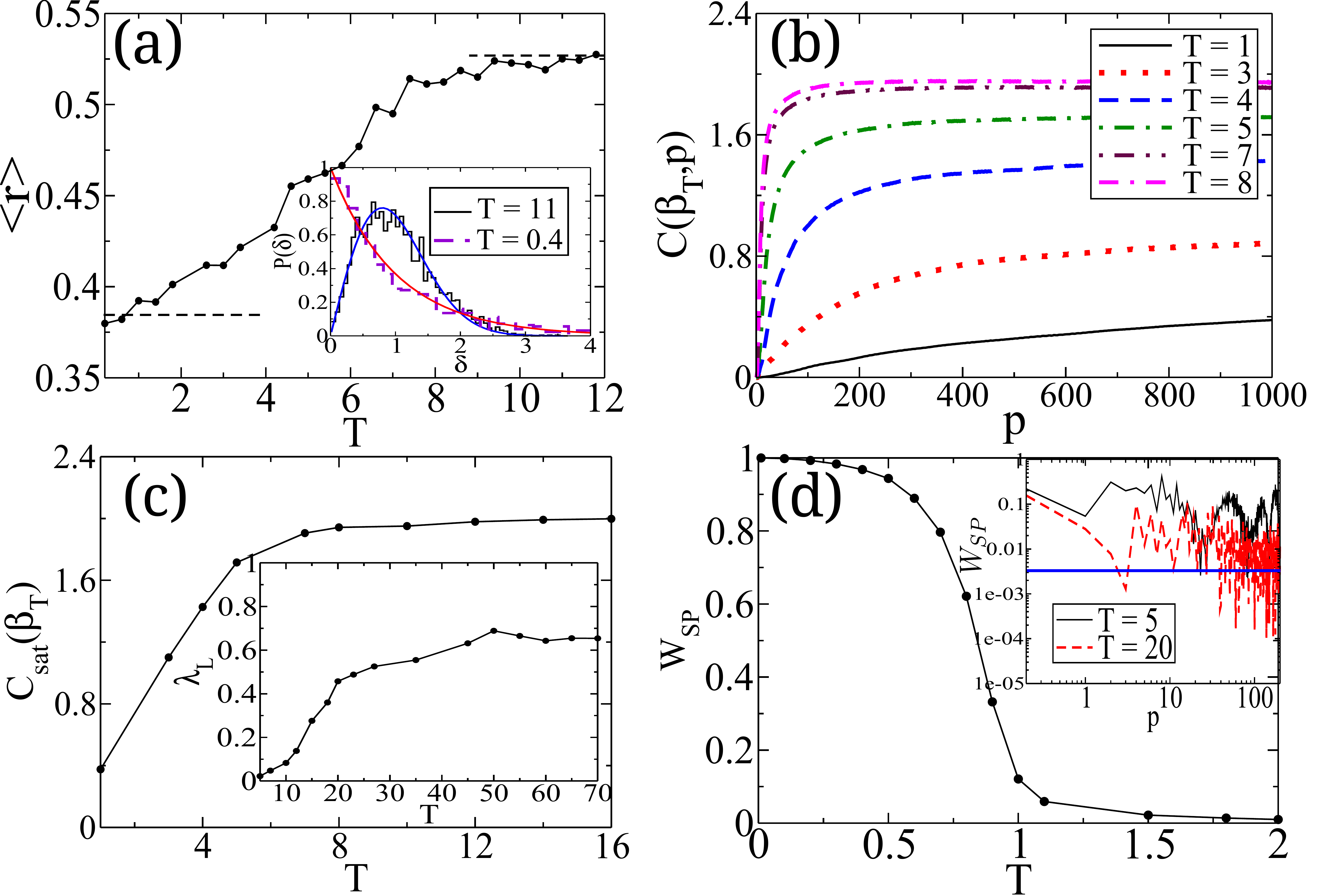}
\caption{(a) $\langle r\rangle$ is shown as function of $T$ and the spacing
distribution of $\delta_{\nu}$'s for two typical values of $T$ given
in the inset; the corresponding probability distributions are shown in
solid curves. (b) $C(\beta_T,p)$ is plotted with number of drive $p$ for
different values of driving time period $T$. (c) Variation of
$C_{sat}(\beta_T)$ is shown as a function of $T$. In the inset,
variation of the growth rate $\lambda_L$ is depicted with $T$.
(d) The survival probability $W_{SP}$ in the steady state
is plotted as function of $T$. The stroboscopic time evolution of $W_{SP}$
is shown in the inset for different $T$. The horizontal line
denotes the GOE value of $W_{SP}$. The other parameters for this
plot are $\lambda = 3$, $\epsilon = 0.47$, $\mathcal{V} = 0.1$ and $\beta_T =
0.1$.} \label{Fig1}
\end{figure}

We first find out the eigenphases, $\phi_{\nu}$ of the corresponding
Floquet operator and order them in $[-\pi,\pi]$. To quantify the
degree of delocalization as well as to identify the
change in the corresponding spectral statistics, we calculate the
ratio between the consecutive level spacing, $r_{\nu} =
\text{min}(\delta _{\nu+1},\delta _{\nu})/\text{max}(\delta
_{\nu+1},\delta _{\nu})$, where $\delta _{\nu} = \phi_{\nu + 1} -
\phi_{\nu}$. We compute the average level spacing ratio $\langle r
\rangle$, which, in the localized regime, $\langle r \rangle \approx
0.386$ signifying that the normalized spacing distribution follows
Poisson statistics, whereas in the delocalized regime $\langle r
\rangle \approx 0.527$ corresponds to the orthogonal class of RMT
\cite{lev_stat,Bogomolny}. From Fig.\ \ref{Fig1}(a) we see that the
value of $\langle r \rangle$ gradually increases from $0.386$ and
reaches a value $0.527$ with the increase in the time period
indicating the thermalization induced by the periodic drive.

Next we investigate the time evolution of the commutator
$C(\beta_T,p)$ constructed from an equivalent local Pauli spin operators
$\hat{W}(\hat{V}) = \hat{\sigma}_z^{l}(\hat{\sigma}_z^{l^{\prime}})$ where
$\hat{\sigma}_z^{l(l^{\prime})} = 2\hat{n}_{l(l^{\prime})} - 1$. We
expect that in the MBL phase $\hat{\sigma}_z^{l}$ commutes with the
Hamiltonian due to the suppression of the kinetic energy resulting
in a very slow growth of $C(\beta_T,p)$. On the other hand the
kinetic energy term becomes significant which can give rise to
non-trivial growth of $C(\beta_T,p)$ in the delocalized regime. In
Fig.\ \ref{Fig1}(b) we have shown the stroboscopic time evolution of
$C(\beta_T,p)$ for different driving time period $T$. We observe
that unlike the large $N$ models, in the delocalized regime the
growth of $C(\beta_T,p)$ is linear and there is no such scrambling
phenomena observed in this driven system. This is apparently due to
the fact that the driving time period is much larger compared to the
typical scrambling time scale; therefore the stroboscopic time
evolution cannot capture this phenomena. However, $C(\beta_T,p)$
saturates eventually in the stroboscopic evolution; the saturation
value $C_{sat}(\beta_T) = \lim_{p \rightarrow \infty} C(\beta_T,p)$
increases with increasing driving time period $T$ and finally for
large $T$ it saturates to the value $C_{sat}(\beta_T) \sim 2$ as
depicted in Fig.\ \ref{Fig1}(c). The relation of the saturation
value of $C_{sat}(\beta_T)$ with that obtained using simple random
matrix models is discussed in the supplementary material
\cite{supp1}. Further we fit the growth of $C(\beta_T,p)$ with a
function $C_{sat}(\beta_T)(1-e^{-p\lambda_{L}})$ where $\lambda_L$
represents the growth rate and increases with increasing $T$ as
illustrated in the inset of Fig.\ \ref{Fig1}(c). We note that both
the behavior of $C_{sat}(\beta_T)$ and $\lambda_L$ resembles the
behavior of $\langle r\rangle$ and therefore can identify the
delocalization crossover with the variation of driving time period
$T$.

\begin{figure}[ht]
\centering
\includegraphics[scale=0.165]{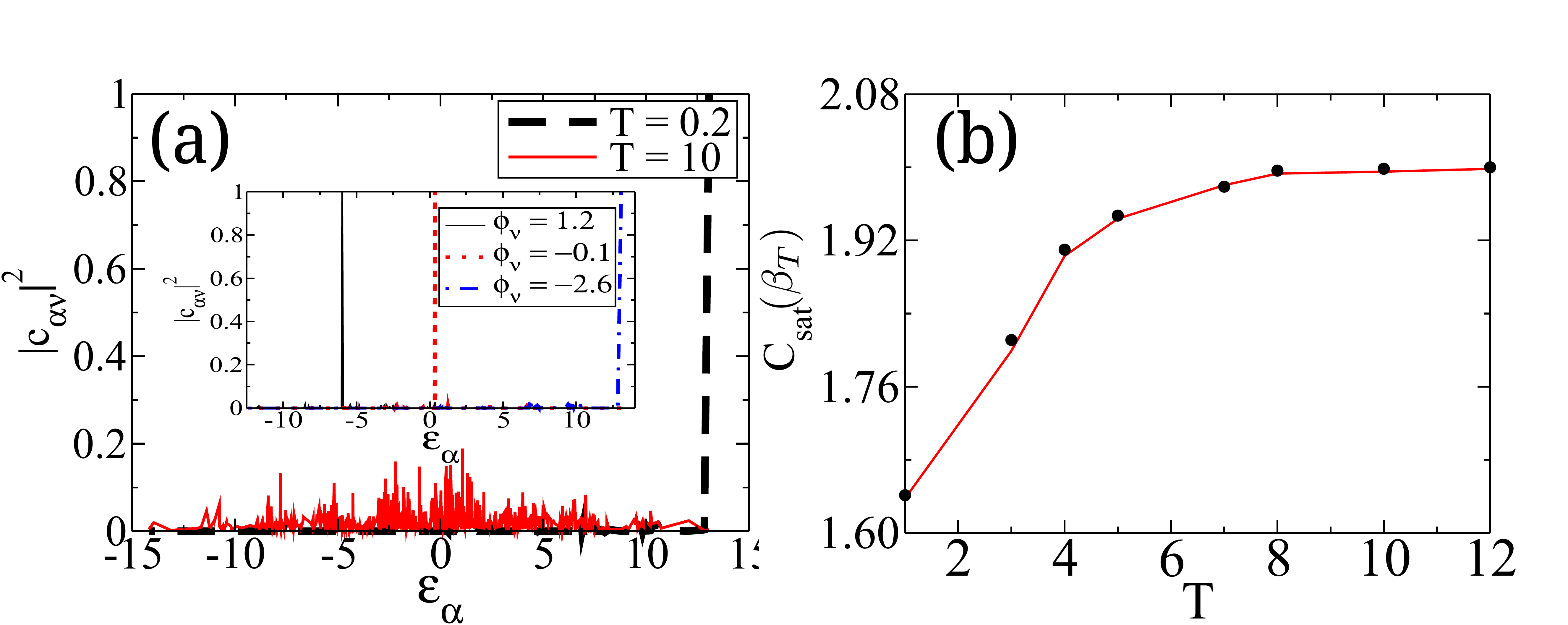}
\caption{(a) The overlap $|c_{\alpha \nu}|^2$ is shown as a function
of $\epsilon_{\alpha}$ for a representative $\phi_{\nu}$
corresponding to the lowest eigenmode of $\mathcal{\hat{F}}$. In the
inset the same has been plotted for $T = 0.2$ and the eigenmodes
corresponding to lowest, middle and upper Floquet band. (b)
$C_{sat}(\beta_T)$ is shown as a function of $T$ for $l =
l^{\prime}$. The solid line is obtained from Eq.\ \ref{OTOC_anal}
and the circles represent the values obtained from full stroboscopic
time evolution. All other parameters are same as in Fig.\
\ref{Fig1}.} \label{Fig2}
\end{figure}

The crossover to delocalized phase can also be captured from the
survival probability \cite{Santos} of an initially prepared state in the coarse of
time evolution which is analogous to `imbalance factor' measured in
the experiments to capture the delocalization transition \cite{Bloch_exp}. In the
dynamical evolution we choose the initial state $|\Psi(0)\rangle$ to
be the ground state of the undriven Hamiltonian. During the
stroboscopic time evolution in presence of drive the survival
probability of the initial state can be computed from $W_{SP}(p) =
|\langle \Psi(0)|\hat{\mathcal{F}}^p|\Psi(0)\rangle|^2$. In the MBL
phase $W_{SP}$ remains close to unity, whereas it decays in the
delocalized regime. We compute the saturation value of $W_{SP}$
in the steady state (obtained for $p\simeq 1000$ in
our numerics) and depict its variation as a function of $T$ in Fig.\
\ref{Fig1}(d). In well inside the delocalized regime $W_{SP}$
saturates to $\sim 3/\mathcal{D}$ [see the inset in Fig.\
\ref{Fig1}(d)] which is in accordance with the RMT prediction
\cite{Santos}, $\mathcal{D}$ being the dimension of the Hilbert
space.

In order to understand the physics behind the decay of the survival
probability, we calculate the overlap of the Floquet states with the
eigenstates of the undriven Hamiltonian. We compute the quantity
$c_{\alpha \nu}=\langle \psi_{\nu} | v_{\alpha} \rangle$, where
$|v_{\alpha} \rangle$ is the eigenstate corresponding to the
$\alpha$th eigenmode of the undriven Hamiltonian. In Fig.\
\ref{Fig2}(a) we have shown $|c_{\alpha \nu}|^2$ corresponding to the
Floquet state with eigenphase $\phi_{\nu}$ as a function of the
eigenenergies $\epsilon_{\alpha}$ of the undriven Hamiltonian. We
observe that in the small $T$ regime typically the Floquet states
have maximal overlap with one of the eigenstates of the undriven
Hamiltonian indicating localization, whereas for higher $T$ the
overlap function $|c_{\alpha \nu}|^2$ spreads over all the eigenmodes
$|v_{\alpha}\rangle$. This observation allows us to consider that in
the localized regime, the local operators $\hat{\sigma}_z^l$ are diagonal in the
Floquet basis. The saturation value of $F$ with $\hat{W} = \hat{V} = \hat{\sigma}_{z}^l$
is independent of $l$ and can be approximated by \cite{supp1},
\begin{equation}
F(\beta_T, p\rightarrow \infty) =
\sum_{\alpha,\nu}\rho_{\beta_T}^{\alpha} |c_{\alpha \nu}|^2 s_{\nu
\nu}^4 \label{OTOC_anal}
\end{equation}
where $\rho_{\beta_T}^{\alpha} = e^{-\beta_T
\epsilon_{\alpha}}/\sum_{\alpha} e^{-\beta_T\epsilon_{\alpha}}$
are elements of the initial thermal density matrix,
$c_{\alpha \nu} = \langle v_{\alpha}|\psi_{\nu} \rangle$ and $s_{\nu
\nu} = \langle \psi_{\nu}|\hat{\sigma}_z|\psi_{\nu}\rangle$. This
approximate analytical formula surprisingly agrees well with the
results obtained from the full stroboscopic dynamics even for large
$T$ as illustrated in Fig.\ \ref{Fig2}(b). From the above expression
it can also be noted that in the delocalized regime the saturation
value $C_{sat}(\beta_T)$ becomes independent of the temperature
scale due to the spreading of the overlap function $|c_{\alpha
\nu}|^2 \sim 1/\mathcal{D}$ [see Fig.\ \ref{Fig2}(a)] \cite{supp1}.
On the contrary in the localized regime (for smaller $T$),
$C_{sat}(\beta_T)$ exhibits a strong temperature dependence and
decreases with increasing $\beta_T$; finally in the delocalized
regime $C_{sat}(\beta_T)$ attains the maximum value $2$ signifying
the infinite temperature thermalization in driven systems
\cite{inf_therm} as illustrated in \cite{supp1}. Therefore this
temperature dependence of the OTOC serve as a indicator of the
degree of localization \cite{Fradkin}

{\it Model II :} In the second case we consider a different type of
drive applied to a system of strongly interacting bosons in a
quasiperiodic potential which is described by the Hamiltonian
\begin{eqnarray}
\hat{H}_0 &=& \sum_{l} \big[ -J
\left(\hat{b}_{l}^{\dagger}\hat{b}_{l+1}+h.c.\right) + \mathcal{V} \hat{n}_l
\hat{n}_{l+1} \label{ham2} \\
&+& \lambda \cos(2\pi \beta l)\hat{n}_{l}\big], \,\, \hat{H}_{1}(t)
= 4\Delta f^{\prime}(\omega t)/T \sum_{l} l \hat{n}_l \nonumber
\end{eqnarray}
where, $\Delta$ is the driving amplitude and $f^{\prime}(x) =
\theta(x-\pi/2) - 2\theta(x-3\pi/2) - \theta(\pi/2-x)$. Such a drive
gives rise to a nontrivial effect on the localization phenomena
which has been explored in Ref.\ \cite{SRay}. In the non-interacting
limit of the above model, it has been shown that there is a domain
of frequency interval within which there appears a delocalized
Floquet band which stems from the underlying chaotic dynamics of the
equivalent classical model \cite{SRay}. This is a counterintuitive
scenario since such a drive in absence of quasiperiodic potential
leads to the suppression of kinetic energy of the time averaged
Hamiltonian \cite{eckardt_rev,arimondo}, hence expected to favor
localization. To explore such phenomena and its connection with the
underlying chaos in the interacting many body system we now follow
the similar procedure as outlined in {\it Model I}.

\begin{figure}[ht]
\centering
\includegraphics[scale=0.158]{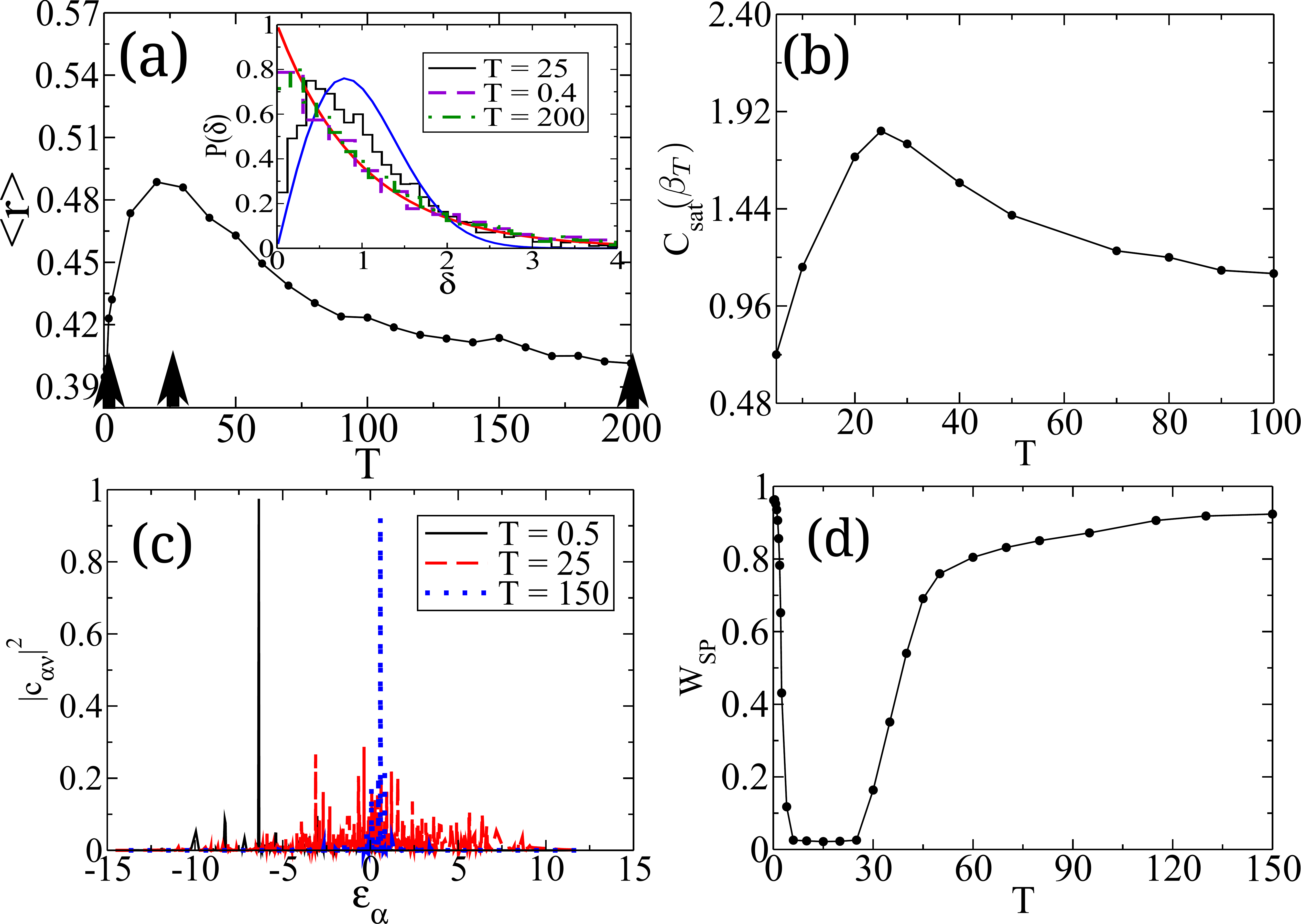}
\caption{(a) Variation of $\langle r\rangle$ is shown as a function
of $T$. In the inset the distribution of the spacing
$\delta_{\nu}$'s are shown for three typical values of $T$ mentioned
therein; the corresponding probability distributions are shown by the solid lines.
$C_{sat}(\beta_T)$ and $W_{SP}$ is plotted with increasing
$T$ in (b) and (d) respectively. (c) The overlap $|c_{\alpha
\nu}|^2$ is shown as a function of $\epsilon_{\alpha}$ for a
representative $\phi_{\nu}$ corresponding to the lowest eigenmode of
$\mathcal{\hat{F}}$. The other parameters are $\lambda = 3$, $\Delta
= 1$, $\mathcal{V} = 0.1$ and $\beta_T = 0.1$.} \label{Fig3}
\end{figure}

First we analyze the Floquet spectrum $\phi_{\nu}$ and compute the
average level spacing ratio $\langle r\rangle$ to characterize the
delocalized as well as the localized phase. In Fig.\ \ref{Fig3}(a)
we have shown $\langle r\rangle$ as a function of driving time
period $T$. In the small $T$ regime as well as for large $T$,
$\langle r \rangle \sim 0.386$ indicating the localized Floquet
states; the corresponding spacing distribution of the eigenphases
exhibit Poisson distribution as shown in the inset of Fig.\
\ref{Fig3}(a). On the other hand, in the intermediate regime
$\langle r \rangle$ increases with increasing $T$ and shows a peak
at $T \sim 25$ exhibiting the level repulsion in the corresponding
spacing distribution as depicted in the inset of Fig.\
\ref{Fig3}(a). To further analyze this non-monotonic behavior of
$\langle r \rangle$, we compute $C_{sat}(\beta_T)$ and plotted it as
a function of $T$ in Fig.\ \ref{Fig3}(b). We see that
$C_{sat}(\beta_T)$ shows a maximum at $T \sim 25$ and decreases on
both the side resembling the non-monotonic behavior of $\langle r
\rangle$. Although the peak values of both the quantities $\langle r
\rangle$ and $C_{sat}(\beta_T)$ are less than that of the GOE limit,
but it clearly distinguishes from the MBL phase and indicates an
approach to thermalization.

To explore the connection of underlying chaos with the
delocalization of the many body Floquet states, we compute the
overlap function $|c_{\alpha \nu}|^2$ and plotted it in Fig.\
\ref{Fig3}(c) for different values of $T$. For small as well as for
larger values of $T$, $|c_{\alpha \nu}|^2$ shows maximum overlap
with one of the eigenmodes of the undriven Hamiltonian in Eq.\ \ref{ham2} indicating
localization, whereas, for intermediate values of $T \sim 25$ the
overlap function spreads over the eigenmodes $|v_{\alpha}\rangle$
showing delocalization of the Floquet states. The peak in
$C_{sat}(\beta_T)$ and the spreading of the overlap function
$|c_{\alpha \nu}|^2$ indicates that such delocalization phenomena
within an intermediate domain of $T$ is a manifestation of the
underlying chaos in the many body system. The delocalization of the
Floquet states further results in the decay of survival probability
$W_{SP}$ around $T \sim 25$. In Fig.\ \ref{Fig3}(d) we have shown
the behavior of $W_{SP}$ as a function of $T$; the dip around $T
\sim 25$ indicates delocalization and the minimum value approaches
the GOE limit. Such a domain of $T$ where delocalized Floquet states
appear can also be captured from the entanglement entropy and has
been illustrated in Ref.\cite{SRay}.

\begin{figure}[ht]
\centering
\includegraphics[scale=0.165]{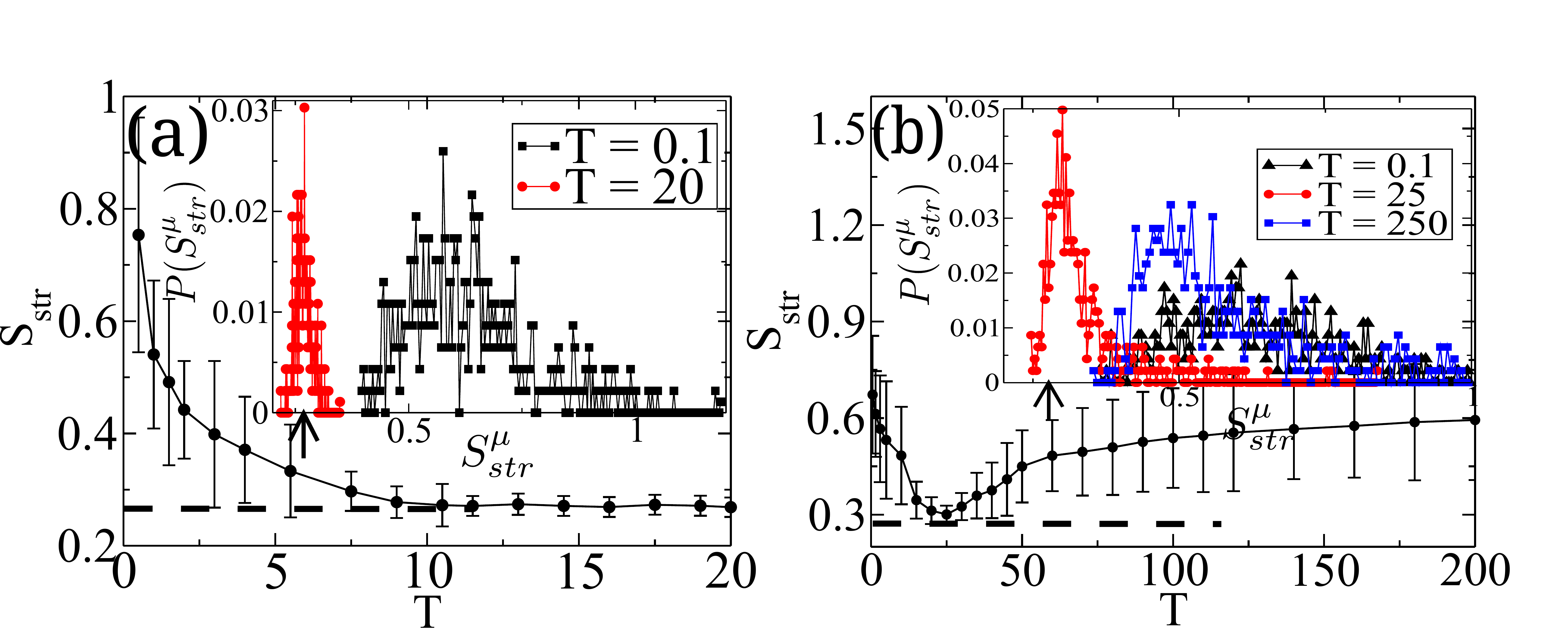}
\caption{$\langle S_{str}\rangle$ is plotted with increasing $T$ in
(a) and (b) corresponding to $Model\, I$ and $II$ respectively. The
error bars indicate the width $\Delta S_{str}$ of the distribution
of structural entropy $P(S^{\mu}_{str})$, shown in the inset of both
the figures for typical values of $T$ mentioned therein. The horizontal
dashed lines indicate the GUE value of the structural entropy.}
\label{Fig4}
\end{figure}

{\it Statistics of OTOC :} Next, we focus on the statistics of OTOC
motivated by a recent observation that the level spacing
distribution of the OTOC corresponding to a single particle chaotic
Hamiltonian exhibits a level repulsion analogous to the Gaussian unitary (GUE)
universality class \cite{Galitski}. In what follows we test the
statistics obtained from the eigenmodes of the OTOC operator
$\hat{F} =
\hat{W}^{\dagger}(p)\hat{V}^{\dagger}(0)\hat{W}(p)\hat{V}(0)$ to
distinguish the delocalization and thermalization phenomena in a
driven many body system. We compute the operator, $\hat{F}$ after
sufficient number of drives and calculate the structural entropy
$S^{\mu}_{str}$ of the $\mu$th eigenmode $|e_{\mu}\rangle$ of
$\hat{F}$, defined as \cite{Verga},
\begin{equation}
S^{\mu}_{str} = -\sum_{\chi}|c_{\chi}^{\mu}|^2 \ln |c_{\chi}^{\mu}|^2 - \ln \xi_{\mu},
\end{equation}
where $c_{\chi}^{\mu} = \langle \chi |e_{\mu}\rangle$ is the overlap
of $|e_{\mu}\rangle$ with the computational basis $|\chi \rangle$
and $\xi_{\mu} = 1/\sum_{\chi}\vert c_{\chi}^{\mu} \vert ^4$ is the
corresponding inverse participation ratio (IPR). For the
eigenvectors of the random matrices of Gaussian unitary class, the
average structural entropy $S_{str}$ approaches to a universal value
$\approx 0.27$ \cite{BM} independent of the dimensionality of the
Hilbert space. In Fig.\ \ref{Fig4}(a),(b) we have shown the
variation of $S_{str}$ with increasing $T$ for both {\it Model I,
II} respectively. The distribution of the structural entropy
$P(S_{str}^{\mu})$ is shown in the inset for different values of $T$
belonging to the localized regime and the delocalized regime. We
notice that in the regime where thermalization occurs in both the
models, structural entropy is sharply peaked at the value $\sim
0.27$ indicating the GUE universality class. On the other hand in
the localized regime, the peak vanishes and $P(S^{\mu}_{str})$ shows
a broad distribution with increasing width $\Delta S_{str} =
\sqrt{\langle S_{str}^{\mu 2}\rangle - \langle
S_{str}^{\mu}\rangle^2}$ shown by the error bars. This observation
confirms that the statistics of OTOC can be an alternate method to
detect the delocalization in a strongly interacting driven system.

{\it Conclusion :} To summarize we have studied the behavior of OTOC
to detect the delocalization transition in a strongly interacting
bosonic system in presence of a quasiperiodic potential subjected to
two types of periodic drives showing distinctly different phenomena.
We have shown that the saturation value $C_{sat}(\beta_T)$, the
survival probability $W_{SP}$, and the spectral property of the OTOC
can efficiently distinguish between the localized and the
delocalized regime. Moreover recent experiments on the measurement
of OTOC in trapped ion and NMR systems \cite{ARey,OTOC_exp} provide
the way to calculate the saturation value $C_{sat}(\beta_T)$ and can
further be implemented to study such delocalization transition in a
system of strongly interacting bosons in driven AA potential. The
survival probability can be measured in experiment from the decay of
an initially prepared state of the system in a similar way as has
been done to measure the `imbalance factor' in the cold atom
experiments \cite{Bloch,Bloch_exp}. It is a general belief that the
growth of $C(\beta_T,p)$ can capture the chaotic behavior of a
quantum system. In previously studied non-interacting AA model under
similar driving protocol it has been shown that the drive induced
delocalization phenomena is connected with the chaotic dynamics of
the corresponding classical model \cite{SRay}. Our present study
also confirms that the delocalization phenomena has a connection
with the underlying chaotic dynamics of an interacting quantum
system even when a direct classical correspondence is absent. Our
results thus provide an alternate approach to diagnose the chaos as
well as its connection with the thermalization in driven interacting
many body systems and can be tested in the experiments in a similar
line of thought as in \cite{ARey,OTOC_exp,Bloch,Bloch_exp}.

\newpage

\begin{widetext}
\appendix

\begin{center}
{\bf SUPPLEMENTAL MATERIAL}
\end{center}

In this supplemental material we provide the results of the full
quantum dynamics governed by a model Hamiltonian consisting of a
mixture of random matrix of Gaussian Orthogonal class (GOE) and a
Poisson matrix. We also discuss the stroboscopic time evolution of
the out-of-time-order correlator (OTOC) as well as discuss the
temperature dependence of its saturation value.

\section{Evolution under GOE matrix}

First, we consider the Hamiltonian describing a
mixed random matrix ensemble given by,

\begin{equation}
\hat{H}_R = \hat{H}_P + \lambda\hat{H}_G/\sqrt{\mathcal{D}}
\end{equation}

where $\hat{H}_P$ is a random banded matrix exhibiting Poisson level
spacing distribution, $\hat{H}_G$ is a GOE matrix and $\mathcal{D}$ is the
size of the matrix. In what follows we construct the Floquet
operator $\mathcal{\hat{F}} = e^{-i\hat{H}_R T}$ and
study its spectral properties as well as the dynamics governed by
$\mathcal{\hat{F}}$. We note that here $\lambda$ is a tuning
parameter and in the limit $\lambda \rightarrow 0$, $\hat{H}_R$ is a
Poisson matrix where as for $\lambda >> 1$, $\hat{H}_R$ resembles a
GOE matrix.

{\it Spectral Statistics :} From the eigenvalue equation
$\mathcal{\hat{F}}|\psi_{\nu}\rangle =
e^{-i\phi_{\nu}}|\psi_{\nu}\rangle$, we first compute the
eigenphases $\phi_{\nu}$ corresponding to the eigenmode
$|\psi_{\nu}\rangle$. We compute the average level spacing ratio
$\langle r \rangle$ as defined in the main text and plotted it as a
function of $\lambda$ in Fig.\ \ref{Fig_supp1}(a). For small
$\lambda$, $\langle r \rangle \sim 0.386$ indicating the Poisson
distribution; further increase in $\lambda$ results in an increase
in $\langle r \rangle$ and finally saturates to $\sim 0.527$
representing the GOE class of the corresponding spacing distribution
as depicted in Fig.\ \ref{Fig_supp1}(a).

{\it OTO Correlator :} To study the time evolution under
$\mathcal{\hat{F}}$, we first construct the initial thermal density
matrix as $\hat{\rho}_{\beta_T} = e^{-\beta_T \hat{H}_P}$ so as to
start from a localized system and evolve it stroboscopically. We
calculate the commutator $C(\beta_T,p)$ and plotted it as a function
of $p$ for different values of $\lambda$ in Fig.\
\ref{Fig_supp1}(b). It can be noted that the growth rate of
$C(\beta_T,p)$ as well as the saturation value $C_{sat}(\beta_T)$
obtained after sufficient number of drives increase with increasing
$\lambda$. This is further illustrated in Fig.\ \ref{Fig_supp1}(c)
where we have shown $C_{sat}(\beta_T)$ as a function of $\lambda$.
It is important to note that in the GOE regime $F(\beta_T,p)$ decays
very fast and becomes vanishingly small leading to $C_{sat}(\beta_T)
\sim 2$ as depicted in Fig.\ \ref{Fig_supp1}(c).

\begin{figure}[ht]
\centering
\includegraphics[scale=0.22]{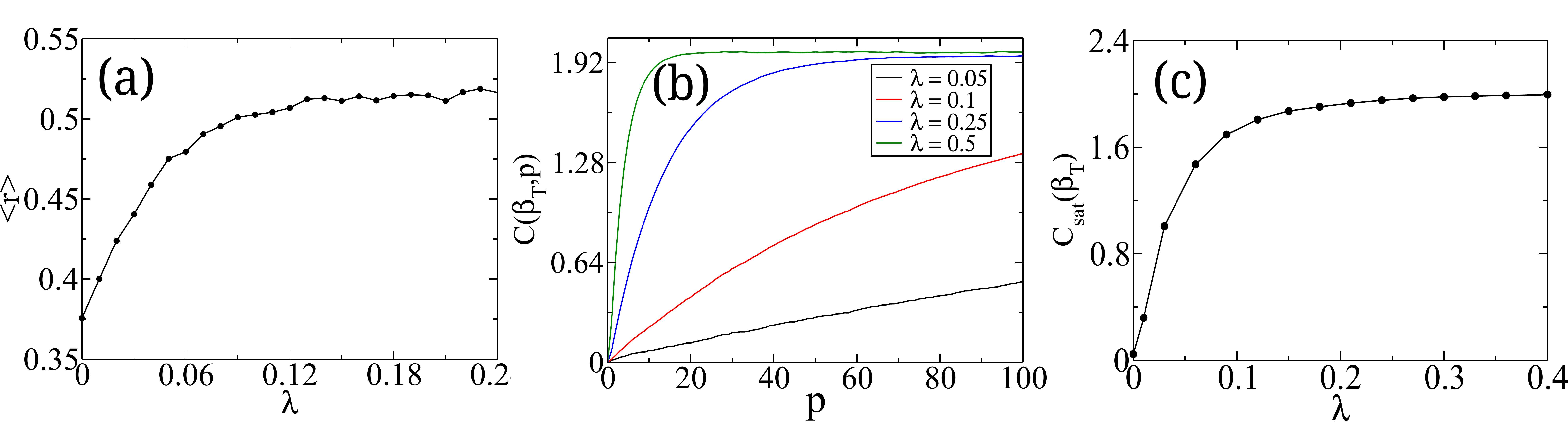}
\caption{(a) $\langle r \rangle$ as a function of $\lambda$. (b)
Stroboscopic time evolution of $C(\beta_T,p)$ for different values
of $\lambda$. (c) Saturation value $C_{sat}(\beta_T)$ is plotted as
a function of $\lambda$.} \label{Fig_supp1}
\end{figure}

\section{Stroboscopic time evolution of OTOC for {\it Model I}}

In this section we will study the stroboscopic evolution of the OTOC
corresponding to {\it Model I} and also study its dependence on
different system sizes in both the MBL phase and in the regime where
thermalization occurs. To this end, we first sketch
the derivation of Eq.(5) of the main text. The quantity $F(\beta_T,
p)$ after $p$ cycles of the drive and at the inverse temperature
$\beta_T$ is given by
\begin{eqnarray}
F(\beta_T, p) &=& {\rm Tr} [ \hat{\rho}_{\beta_T} \hat{W}(p) \hat{V}(0) \hat{W}(p) \hat{V}(0)] \nonumber\\
&=& \sum_{\alpha} e^{-\beta_T \epsilon_{\alpha}} \langle v_{\alpha}|
{\hat{\mathcal F}}^{\dagger} \hat{W}(0) {\hat{\mathcal F}}\, \hat{V}(0)\,  {\hat{\mathcal
F}}^{\dagger} \hat{W}(0) {\hat{\mathcal F}} \hat{V}(0)|v_{\alpha}\rangle
/Z\label{otocan1}
\end{eqnarray}
where ${\hat{\mathcal F}}$ is the evolution operator and $|v_\alpha\rangle$,
$\epsilon_{\alpha}$ denotes eigenvectors and eigenvalues of $\hat{H}(t=0)$ (see Eq.\ 4 in the main text)
and $Z= \sum_{\alpha} \exp(-\beta_T \epsilon_{\alpha})$ is the
corresponding partition function. We then decompose $F$ in terms of
its eigenstates $|\psi_{\nu} \rangle$ and eigenenergies $\phi_{\nu}$ (as
defined in the main text. A few lines of algebra yields
\begin{eqnarray}
F(\beta_T, p) &=&  \frac{1}{Z} \sum_{\alpha} \sum_{\mu, \nu,\lambda,
\mu' \nu'} c_{\alpha \mu}^{\ast} c_{\nu \alpha} e^{i(\phi_{\mu} +
\phi_{\lambda}-\phi_{\mu'} -\phi_{\nu'})p T}\, W_{\mu \mu'}(0)
W_{\lambda \nu'}(0) V_{\mu' \lambda}(0) V_{\nu' \nu}(0).
\label{otocan2}
\end{eqnarray}
Here $c_{\alpha \nu} = \langle v_{\alpha} |\psi_{\nu} \rangle$ denotes the
overlap function, $T$ is the drive time period,
and ${\mathcal F}$ and $W_{\nu \nu'}(0)$ denotes the matrix element
of $\hat{W}$ between Floquet eigenstates $|\psi_{\nu}\rangle$ and $|\psi_{\nu'}\rangle$.
For $p \to \infty$, the contribution to $F$ is obtained from terms
for which $\phi_{\mu} + \phi_{\lambda} = \phi_{\mu'} + \phi_{\nu'}$.
Furthermore, numerically, we find that for $\hat{W} = \hat{V} = \hat{\sigma}_z$, these
matrix elements have maximal contribution from the diagonal terms.
Denoting $\langle \psi_{\nu}| \hat{\sigma}_z |\psi_{\nu'}\rangle \simeq s_{\nu \nu'} \delta_{\nu
\nu'}$, we finally obtain,
\begin{equation}
F(\beta_T, p\rightarrow \infty) =
\frac{\sum_{\alpha,\nu} e^{-\beta_T \epsilon_{\alpha}} |c_{\alpha
\nu}|^2 s_{\nu \nu}^4 }{\sum_{\alpha} e^{-\beta_T
\epsilon_{\alpha}}} \label{OTOC_anal}
\end{equation}
which is Eq. (5) of the main text. We find that in the delocalized
regime $|c_{\alpha \nu}|^2 \sim 1/\mathcal{D}$ where $\mathcal{D}$
is the dimension of the matrix. This results in the saturation value
of $C_{sat}(\beta_T) \sim 2$ indicating that such a driven system
thermalizes to infinite temperature where $C_{sat}(\beta_T)$ becomes
independent of $\beta_T$.

In Fig.\ 1(a) of the main text it has been shown that the growth
rate of $C(\beta_T,p)$ is very small in the MBL phase, on the other
hand $C(\beta_T,p)$ grows very fast in the delocalized regime. In
contrast the OTOC, $F(\beta_T,p)$ shows a very slow power law decay
in the MBL phase as depicted in a logarithmic plot in Fig.\
\ref{Fig_supp2}(b) and shows a very strong system size dependence in
the saturation value. On the other hand, for large driving time
period $T$ OTOC decays exponentially fast and saturates to a
vanishingly small value as shown in Fig.\ \ref{Fig_supp2}(c).

\begin{figure}[ht]
\centering
\includegraphics[scale=0.22]{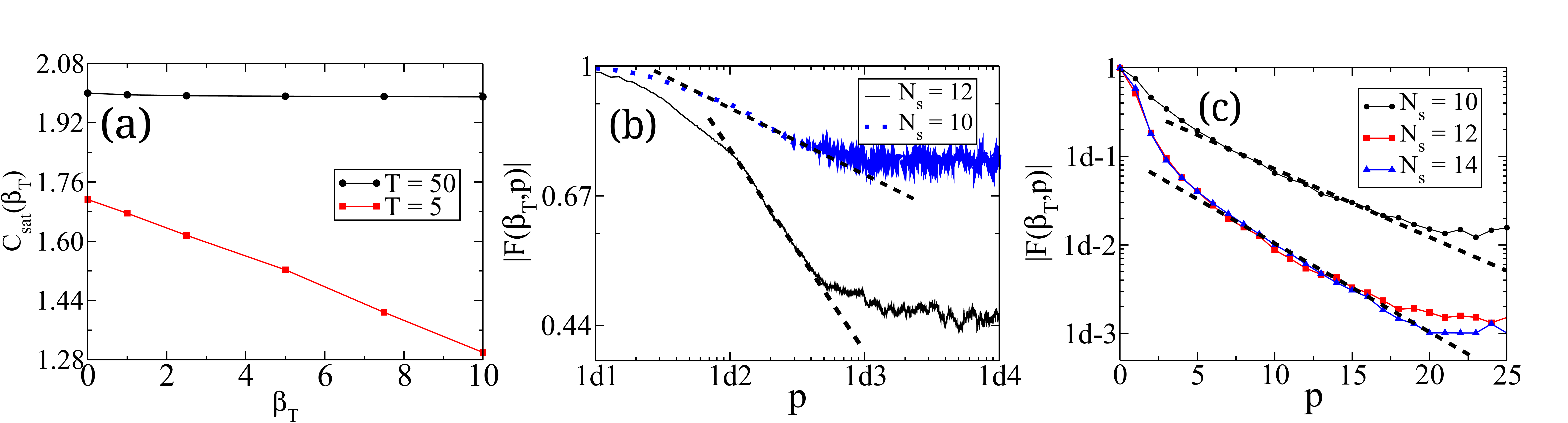}
\caption{(a) $C_{sat}(\beta_T)$ is shown as a function of $\beta_T$.
The stroboscopic time evolution of $|F(\beta_T,p)|$ is shown in the
localized regime for $T = 5$ in (b) and in the delocalized regime
for $T = 50$ in (c). The dashed lines in (b) and (c) indicate the linear
behavior in log-log plot and in semi-log plot respectively.} \label{Fig_supp2}
\end{figure}

{\it Temperature dependence :} Further we study the temperature
dependence of the OTOC. First we note that for large $T$ when the
system thermalizes, there is no such temperature dependence and
$F(\beta_T,p)$ decays exponentially resulting in the saturation
value $C_{sat}(\beta_T) \sim 2$. Whereas for small $T$ i.e. in the
MBL phase, the saturation value of the OTOC increases with
decreasing temperature resulting in the decrease in
$C_{sat}(\beta_T)$ with increasing $\beta_T$ as illustrated in Fig.\
\ref{Fig_supp2}(a). Such a growth of OTOC can be
understood as follows. We first note that in Eq.\ \ref{OTOC_anal},
$\sum_{\nu} |c_{\alpha \nu}|^2 =1$ for all $\alpha$; moreover, it
can be checked numerically $s_{\nu \nu}^4 <1$ for all $\nu$. Thus
the quantity ${\mathcal L}_{\alpha} = \sum_{\nu} |c_{\alpha \nu}|^2
s_{\nu \nu}^4 <1$, which ensures the convergence of the numerator of
Eq.\ \ref{OTOC_anal} since $\epsilon_{\alpha}$ can always chosen to
be positive without any loss of generality. In terms of ${\mathcal
L}_{\alpha}$, one can write
\begin{eqnarray}
F(\beta_T, p\rightarrow \infty) &=&  \frac{\sum_{\alpha} e^{-\beta_T
\epsilon_{\alpha}}{\mathcal L}_{\alpha }}{{\sum_{\alpha} e^{-\beta_T
\epsilon_{\alpha}}}} = \frac{N}{Z} \label{OTOC_anal2}
\end{eqnarray}
Using this expression, it is straightforward to see, that
\begin{eqnarray}
\frac{\partial F(\beta_T, p\rightarrow \infty)}{\partial \beta_T}
&=& \frac{1}{Z^2} \sum_{\alpha \ne \alpha'}
e^{-\beta_T(\epsilon_{\alpha} + \epsilon_{\alpha'})}
\epsilon_{\alpha} {\mathcal L}_{\alpha'} \, >0
\end{eqnarray}
Thus $F(\beta_T, p\rightarrow \infty)$ must increase with increasing
$\beta_T$ leading to decrease of $C_{sat}(\beta_T)$ with $\beta_T$
as shown in Fig.\ \ref{Fig_supp2}(a).

\end{widetext}

\end{document}